# Numerical and experimental analysis of the birefringence of large air fraction slightly unsymmetrical holey fibres

Laurent LABONTE[*], Dominique PAGNOUX[*], Philippe ROY[*],

Faouzi BALHOUL[**], Mourad ZGHAL[**]

* Institut de Recherche en Communications Optiques et Microondes (IRCOM), Unité Mixte CNRS-Université de Limoges n°6615, 123 Avenue A. Thomas, 87060 Limoges Cedex, France
** Ecole Nationale d'Ingénieurs de Tunis (ENIT), BP37, 1002 Le Belvédère, Tunis, Tunisia

*corresponding author : Laurent LABONTE*
*email   : labonte@ircom.unilim.fr         tel :  +33 5 55 45 74 17     fax   +33 5 55 45 72 53*

**Abstract :** Careful numerical computations show that very slight geometrical imperfections of the cross section of actual large air fraction holey fibres ($d/\Lambda > 0.6$) may induce surprisingly high birefringence, corresponding to beat lengths as short as few millimetres. The spectral variations of this birefringence obeys laws similar to those of elliptical core Hi-Bi holey fibres with low air fraction. For all the tested fibres, the group birefringence numerically deduced from the only shape birefringence is in good agreement with the measured one that does not varies when strongly heating the fibres. These computations and measurements show that the contribution of possible inner stress to the birefringence is negligible.

**Keywords :** microstructured fibres, finite element method, shape birefringence, fibre characterisation

## I. Introduction:

Thanks to their special propagation properties, microstructured optical fibres (MOFs) have attracted a considerable interest in recent years. MOFs with a solid core operating following the total internal reflection principle are commonly known as holey fibres (HFs). The stack and draw technique operated for manufacturing HFs opens a large range of opportunities for tailoring the propagation characteristics such as the spectral single mode range of operation, the effective area of the fundamental mode or its chromatic dispersion versus the wavelength. For a large number of applications, in particular telecommunications and non linear optics, a particular attention must be paid to the birefringence of HFs. Most of them are based on a triangular arrangement of air holes around a pure silica core, providing a $\pi/3$ symmetry of the refractive index profile of the structure. In such perfect single mode fibres, this symmetry of order greater than 2 results in the degeneracy of the two orthogonally polarized modes supported by the fibre [1].

Into actual HFs, the degeneracy may easily be broken due to anisotropic stress into the structure (stress birefringence) and, above all, to distortions in the geometry of the cross section of the fibre (shape birefringence). Highly birefringent (Hi-Bi) HFs can be designed by deliberately breaking the $\pi/3$ symmetry of the triangular lattice of holes [2,3]. Numerous recent papers have already been devoted to the study of the non intentional shape birefringence into HFs, either theoretically [4] or experimentally [5]. They generally concern HFs having a ratio hole diameter/pitch (d/Λ) lower than 0.6. However, for many applications requiring a strong confinement of light (non linear optics such as supercontinuum generation or soliton applications) small core high numerical aperture fibres are needed. In this paper, we study the shape birefringence and the group birefringence of different very large air-fraction actual HFs (d/Λ>0.6), with very slight difference in geometry compared to the ideal perfectly symmetrical one. On the one hand, the shape birefringence is computed versus the wavelength by means of a vectorial Finite Element Method (FEM) applied to the actual cross section of the considered fibre. The curve is fitted with an empirical law and the geometrical group birefringence is deduced. On the other hand, the actual group birefringence is measured by means of a modulated spectrum analysis technique. The numerical and experimental results are compared and discussed. The effects of slight imperfections of the geometry of the cross section on the birefringence are pointed out.

**II. Numerical computation of the shape birefringence and of the corresponding group birefringence**

   **II. 1 Modelling method**

Four different HFs with a large air-fraction have been manufactured using the stack and draw technique at IRCOM (fibre 1-3) and at Alcatel Research and Innovation Center (fibre 4). Their geometrical characteristics d and Λ are reported in the table I -Part1, together with a Scanning Electron Microscope (SEM) picture of their cross section. Because the d/Λ ratio is significantly higher than 0.4, these fibres cannot be considered as endlessly single mode [6]. However, we have verified using an azimuthal far-field analysis technique, that they are single mode over the [1530-1560nm] bandwidth of the source used in our characterisation device [7]. In the following, the 1540nm medium wavelength of this source is considered as the operating wavelength.

Many vectorial methods have already been proposed to compute the propagation constants and the electric or magnetic field distribution of the modes into HFs. However, most of them only apply either to perfectly symmetrical structures or to fibres with circular and/or elliptical holes [8,9]. To model as accurately as possible the actual cross section of a manufactured HF, we apply a vectorial FEM to a structure deduced from the SEM picture of the fibre. On this picture, the air holes appear in dark and the silica in pale (figure 1a). By means of a numerical treatment that consists in comparing the darkness of each point of the SEM picture to a decision level determined in the grey-scale, we locate the silica and the air regions and thus the holes boundaries. The one-colour corresponding figure, on which the boundaries of the holes are plotted, is called "the mask" (figure 1b). Each hole and the surrounding silica region are then split in a mesh of triangular elementary subspaces, with dimensions small enough for properly describing each domain (figure 1c) [10]. The effective indexes of the two orthogonal polarisation modes, respectively $ne_x$ and $ne_y$, are computed by solving the Maxwell equations at each node of the grid and the shape birefringence B is simply deduced with :

$$B = |ne_x - ne_y| \quad (1)$$

Because the considered shape birefringence is due to only slight geometrical imperfections of the cross section, the calculation must be achieved very carefully in order to provide reliable values. As mentioned above, for a better description of the cross section of the fibre, the size of the triangular elementary subspaces of the mesh must be as small as possible. However, as the size is decreased, the number of necessary subspaces is increased, inducing an increase of both the necessary memory and the computation time. Typically, these dimensions are chosen to be equal to $\lambda/10$ in the regions where the electric field is expected to be high, i.e. in from the centre to the first ring of holes, and it is equal to $\lambda/7$ elsewhere. For all the tested fibres, we have verified that still reducing these dimensions does not induce any significant change of the obtained effective index values.

Let us note that, even when modelling perfectly six-fold symmetrical HFs with the FEM method, the mesh has not the same $\pi/3$ angular symmetry as the fibre has, as shown in the example of the typical mesh of one circular 1.4µm diameter hole (figure 1d). Moreover, the silica-air boundaries are not perfectly circular but are described by a polygon with a large number of edges. The rotational symmetry of the whole structure is then broken and a residual numerical birefringence (RNB) is found. As it can be expected, this RNB increases as the interaction between the polarisation modes and the holes increases, because of a higher influence of the non symmetry of the mesh at the hole boundaries. This means that the RNB increases as the ratio $\Lambda/\lambda$ decreases and as the ratio $d/\Lambda$ increases. The RNB of the 5 studied fibres has been evaluated considering virtual perfectly symmetrical HFs which parameters d and $\Lambda$ are the mean ones of the corresponding actual fibres (table I -part 2). At $\lambda=1540$nm, the RNB has been found lower than 3 $10^{-6}$ except for fibre 5 for which it reaches 8 $10^{-6}$, corresponding to a beat length $L_b=\lambda/B$ of about 20cm. This larger RNB is due to the higher $d/\Lambda$ value. These RNB values are higher than those obtained by a plane wave expansion method reported in [4]. However, no conclusion on the relative accuracy of the methods can abruptly be drawn because we consider fibres with significantly higher $d/\Lambda$ values (0.6-0.9 instead of 0.48) and lower $\Lambda/\lambda$ values (1.3-1.6 instead of 2.-10.). As we will see further, the RNB remains significantly lower than the computed birefringence of the actual considered fibres.

In order to obtain a mask giving a proper description of the actual cross section, it is necessary to avoid any parallax error when taking the picture and to choose the proper decision level when operating its numerical treatment. Using the FEM software, we have investigated the influence of a realistic change in this decision level on the computed birefringence. We evaluate the maximum uncertainty due to the numerical treatment of the picture to $\pm 5\%$.

**II. 2 Numerical results**

The phase birefringence of each fibre has been computed versus the wavelength from 900nm to 1700nm (figure 2). At λ=1540nm, it is comprised between $6.5\ 10^{-4}$ and $10\ 10^{-4}$ for all the tested fibres, except for fibre 2. The shape birefringence of this fibre is somewhat lower ($1.24\ 10^{-4}$) in spite of a stronger confinement of the polarisation modes, suggesting that the geometrical imperfections of the structure are smaller. Let us note that the reliability of the computation does not suffer from the RNB because, even in the worse case, the RNB is at least 100 times lower than B and remains negligible. Whatever the fibre, the beat length is of the order of few millimeters at any wavelength, so that the studied HFs can be considered as Hi-Bi fibres. These values are similar to those reported in [11], concerning an elliptical shaped core HiBi-HF. This last fibre was characterised by a large dimensional ratio in the elliptical core (2:1). Compared to fibres 1-4, it had lower d/Λ ratio (d/Λ=0.59), significantly larger pitch (Λ=7.4µm) and hole diameter (d=4.4µm). The high geometrical birefringence of fibres 1-4 must be attributed to the considerable influence of the geometrical imperfections of the structures on the field distributions because of the tight confinement into these small-core large-air-fraction fibres.

The geometrical birefringence is the consequence of the combination of the imperfections related to each hole, i.e. mainly the difference between their actual position and/or size and the ideal one, or a possible distortion of their shape. Because of their weakness, these imperfections are very difficult to identify and cannot be precisely measured. In order to evaluate an order of magnitude for these imperfections, the ideal perfectly symmetrical cross section of each of the considered actual fibres as close as possible to the corresponding SEM picture, has been drawn with a computer assisted drawing software. A careful superimposition of the ideal cross section and of the corresponding SEM picture shows that the shift of any hole in any direction remains lower than 20 nm. The possible size discrepancy between holes from the same ring is lower than ± 3% . No significant shape discrepancy between holes from the same ring is revealed.

Let us note that it should be very hazardous to compare the extent of the geometrical imperfections of different fibres by only considering their birefringence, because this last parameter is also very sensitive to the mean optogeometrical parameters of each fibre.

To tentatively compare the extent of the geometrical imperfections into the different actual fibres, we associate to each one an " equivalent birefringent HF (EBHF)" having the same birefringence as the considered one at a given operating wavelength. This EBHF consists in a perfectly symmetrical fibre with the mean d and Λ parameters of the actual one, suffering from a single defect that brings about the same birefringence as that of the actual considered one. As a kind of defect, we have arbitrarily chosen a symmetrical diametral shift ΔS of two opposite holes in the first ring of the perfect fibre. Symmetrical shifts towards the periphery and towards the centre are respectively noted $\Delta S_+$ and $\Delta S_-$ (figure 3). Thus, for a given actual fibre, the higher the birefringence is, the larger ΔS is. But for a given amount of birefringence, ΔS obviously depends on the opto geometrical parameters of the considered actual fibre. In Table I part 4, we have reported both $\Delta S_+$ and $\Delta S_-$ corresponding to the four tested fibres, the operating wavelength being equal to 1540nm. As the confinement of the field is stronger when the shift is operated towards the centre, $\Delta S_-$ is systematically lower than $\Delta S_+$. $\Delta S_+$ is found to be equal to about Λ/10 for fibres 1 and 3, Λ/20 for fibre 4 and Λ/80 for fibre 2. Let us note that, even if the shape birefringence of the fibre 4 is as high as that of fibres 1 and 3, the associated $\Delta S_+$ is significantly lower because of the stronger confinement of the field (higher d/Λ) into this fibre. As expected, these values are very small compared to the pitch, and this confirms that very weak imperfections are sufficient for inducing high birefringence in large air fraction HFs. This points out that, for fabricating low birefringence large air fraction HFs, a particular attention must be paid to the control of the manufacturing process.

The shape birefringence of the EBHFs which geometrical characteristics are reported in the part 4 of table I is plotted versus the wavelength from 900nm to 1700nm, on figure 4. The superimposition with the shape birefringence of the corresponding actual fibre shows a surprisingly good agreement over the whole range of wavelengths, for the four fibres. These results suggest that knowing the shape birefringence of an actual slightly unsymmetrical HF at a given wavelength, the associated EBHF can be used for predicting the shape birefringence at any wavelength.

The shape birefringence of the tested fibres dramatically increases with the wavelength due to a larger interaction with the air holes (figure 2). As it is commonly supported in the literature for this kind of fibres, we make the assumption that the phase birefringence is a function of the wavelength on the form [2] :

$$B = \alpha \cdot \lambda^k \quad (2)$$

where $\alpha$ and k are two constants depending on the fibre. For each studied fibre, the spectral curve of B has been fitted by such a function, as plotted on figure 2. We can observe a very good agreement between the fit and the value of B computed with the FEM over the whole range of wavelengths. The uncertainty on k induced by the ±5%uncertainty on B is very slight and can be neglected. The value of k for fibres 1, 3 and 4 is close to 2.5 (2.46<k<2.59) but it appears somewhat higher for fibre 2 (k=3.05). This is a direct consequence of the stronger confinement of the polarisation modes into this fibre. The obtained values of k are comparable to the 2.58 value reported in [11], confirming that the fibres under test exhibit the same behaviour as elliptical core Hi-Bi HFs.

The group birefringence $B_g$ due to the geometrical imperfections of the fibre is related to B by [11] :

$$B_g = B - \lambda \frac{dB}{d\lambda} \quad (3)$$

Using relation (2), $B_g$ can be expressed as :

$$B_g = B.(1-k) \quad (4)$$

The relation (4) shows that the phase and group birefringence of the tested fibres are opposite in sign, i.e. light polarised following the fast polarisation axis has a group velocity lower than light polarised following the slow one.

The absolute value of $B_g$ is deduced from the previously computed B and k values, with obviously, the same uncertainty as B (approximately ±5%). On the figure 5, we have plotted $B_g$ versus the wavelength, for the four tested fibres. These curves show a large increase of $B_g$ with the wavelength, since it is multiplied by more than 2 when the wavelength is increased from 1100nm to 1550nm. The group birefringence due to geometrical imperfection is very similar in fibres 1, 3 and 4, in spite of significantly different optogeometrical parameters. The spectral behaviour of $B_g$ is notably different in fibre 2, because of its lower shape birefringence and higher k : $B_g$ is about 10 times lower at short wavelengths ($\lambda$=900nm) and remains only 5 times lower at longer wavelengths (($\lambda$=1550nm).

In the following section, we compare the numerical values of $B_g$ with the group birefringence $B_G$ measured around 1540nm, and we discuss the obtained results.

**III. Measurement results and discussion**

The phase birefringence of an optical fibre is generally deduced from the measurement of the beat length $L_b$. Different methods, such as the measurement of the polarised transversally diffused signal [12], the polarisation time domain reflectometry method [13] or the magneto-optical method [14,15], are available. However, they are either very difficult to implement [12] or they cannot apply for HiBi fibres with $L_b$ as short as few mm [13-15]. We have measured the group birefringence of the tested fibres by means of the set-up shown on figure 6a, based on the well-known spectral analysis technique. The polarized light from a broadband source (Keopsis double clad erbium amplifier; ASE in the range 1530-1560nm) is launched at 45° to the neutral axis of a sample of the fibre under test (length L≈2m) in order to equally excite the two polarisation modes. At the output end, two equal parts of each polarisation mode interfere through an analyser which transmission axis is set at 45° to the neutral axis of the fibre. The spectrum of the detected power downstream the analyser exhibits a ripple with a non constant period $\delta\lambda$ related to $B_G$ by :

$$B_G = -\frac{\lambda_0^2}{L.\delta\lambda}$$ (4) where $\lambda_0$ is the wavelength in the centre of the considered period.

The figure 6b is a typical interferogram measured by means of an optical spectrum analyser *(Anritsu MS9702A resolution 0.1nm)*. The group birefringence $B_G$ measured at $\lambda$=1540nm, for fibres 1 to 4 is equal to $1.17\ 10^{-3}$, $2.93\ 10^{-4}$, $1.39\ 10^{-3}$ and $1.19\ 10^{-3}$ respectively, with an uncertainty evaluated to about ±5%. On figure 5, we have plotted this value for each fibre. Concerning fibres 1,3 and 4 it is in very good agreement with the computed $B_g$, since the discrepancy remains lower than 10% in each case. Concerning the fibre 2, the discrepancy reaches 20%. However, the value of $B_g$ strongly depends on the fitting function. With a fit of B by the usual function (2) over the range of wavelengths restricted to [1400nm-1700nm], the values of $\alpha$ and k are slightly corrected ($\alpha$=1.184 $10^{-14}$ and k=3.1411 respectively) and the corresponding value of $B_g$ at 1540nm is only 11% higher than the measured $B_G$. At least, $B(\lambda)$ can be properly fitted with a polynom of order 3 over the restricted range of wavelengths [1400nm-1700nm]. In this case, the discrepancy between the value of $B_g$ calculated with the relation (3) at 1540nm and the measured $B_G$ is lower than 8%.

The above comparison shows that the group birefringence due to geometrical imperfections of the tested fibres is very close to the measured group birefringence, for all the tested fibres. This suggests that the contribution of the possible residual inner stress to the birefringence of the fibres is very weak. To study thoroughly this question, a thermal treatment from 20°C to 1300°C has been applied to the fibres in order to relax possible stress. As expected, no measurable change on $B_G$ has been observed. This confirms that the surprisingly high birefringence of the tested large air fraction HFs can be quite exclusively attributed to their slight geometrical imperfections.

Let us notice that all along this study the cross section of the fibres has been implicitly assumed to be unchanged along the z axis. This assumption remains realistic considering that the tested samples are fairly short (~2m) and that they have been taken on a reel far from the extremity, at a location where the manufacturing parameters such as the furnace temperature, the pressure and the drawing speed were well stabilized. However, due to the large influence of very slight imperfections on the birefringence, it should be of great interest to investigate the possible fluctuations of the birefringence all along the fibre.

## IV. Conclusion

In this paper, we report the very first both numerical and experimental analysis of the shape birefringence and the group birefringence of actual holey fibres with a large air fraction into the cladding (d/Λ>0.6). Four different large air fraction single mode holey fibres have been characterised. We have first shown that their shape birefringence can reliably be computed by means of a vectorial finite element method, using a mask provided by a proper numerical treatment of the SEM picture of the cross section.

In spite of their apparent pretty good symmetry, these fibres suffer from surprisingly high shape birefringence (from $1.2\ 10^{-4}$ to $10^{-3}$ at 1550nm), due to slight geometrical imperfections of their cross section. We have shown that this amount of birefringence, corresponding to beat lengths as short as few millimeters, is the same as that induced into perfectly symmetrical holey fibres by a very small diametrical shift (much shorter than Λ/10) applied to two opposite holes from the first ring. We have also pointed out that the spectral variations of the birefringence in the fibres under test are very similar to those of elliptical core Hi-Bi HFs with lower air fraction. The group birefringence numerically deduced from the only shape birefringence is in very good agreement with the measured one for all the fibres. Furthermore, no change of the group birefringence has been measured after heating the fibres from 20°C to 1300°C for relaxing possible stress. These last computations and measurements suggest that the contribution of possible inner stress to the birefringence of the fibres is negligible.

In conclusion, we have shown that the surprisingly high birefringence into large air fraction holey fibres is due to very slight geometrical imperfections of the cross section. As a consequence, the fabrication of low birefringence large air fraction HFs can be achieved only if the geometrical dissymmetry of the cross section is drastically limited, requiring a particularly careful control of the manufacturing process.

Acknowledgements : The authors are grateful to Jean-Louis Auguste, Pierre-Olivier Martin and Jean-Marc Blondy from IRCOM for their unvaluable contribution in manufacturing the fibres 1 to 3, and to the Alcatel Research and Innovation Center for providing samples of fibre 4.

|  |  | Fibre 1 | Fibre 2 | Fibre 3 | Fibre 4 |
|---|---|---|---|---|---|
| Part 1 : Geometrical characteristics | Cross section (SEM Picture) | 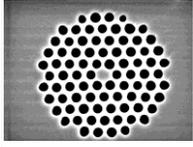 | 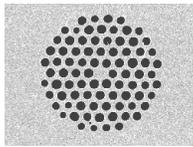 | 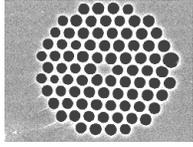 | 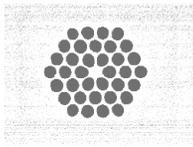 |
|  | d (µm) | 1.46 | 1.4 | 1.8 | 2.2 |
|  | Λ(µm) | 2.15 | 2. | 2.26 | 2.4 |
|  | d/Λ | 0.68 | 0.7 | 0.8 | 0.92 |
| Part 2 : Shape birefringence | Numerical biref @ 1540nm | $3.2 \times 10^{-6}$ | $3.7 \times 10^{-6}$ | $1.2 \times 10^{-6}$ | $8 \times 10^{-6}$ |
|  | Computed B @ 1540nm | $6.87 \times 10^{-4}$ | $1.22 \times 10^{-4}$ | $9.64 \times 10^{-4}$ | $7.66 \times 10^{-4}$ |
|  | α (with λ in nm) | $4.792 \times 10^{-12}$ | $2.334 \times 10^{-14}$ | $5.380 \times 10^{-12}$ | $1.058 \times 10^{-11}$ |
|  | k | 2.559 | 3.049 | 2.589 | 2.466 |
| Part 3: Group birefringence | Computed $B_g$ @ 1540nm | $1.07 \times 10^{-3}$ | $2.50 \times 10^{-4}$ | $1.53 \times 10^{-3}$ | $1.12 \times 10^{-3}$ |
|  | Measured $B_G$ @ 1540nm | $1.17 \times 10^{-3}$ | $2.93 \times 10^{-4}$ | $1.41 \times 10^{-3}$ | $1.19 \times 10^{-3}$ |
| Part 4 : Equivalent Birefringence Holey Fibre | $\Delta S_+$ @ 1540nm (nm) | 227 | 25 | 220 | 120 |
|  | $\Delta S_-$ @ 1540nm (nm) | 153 | 21 | 150 | 96 |

Table I : measured and computed parameters, related to the birefringence of the four tested fibres

**Figure captions :**

Figure 1 : SEM picture of the cross section of fibre1 (a), mask (b), mesh of one hole of this fibre (c), mesh of a circular hole of a perfectly symmetrical holey fibre (d)

Figure 2 : computed shape birefringence of the four tested fibres versus wavelength and corresponding fitting curve on the form $\alpha.\lambda^k$

Figure 3 : cross section of the equivalent birefringence holey fibre (EBHF), consisting in a perfect HF with a symmetrical diametrical shift $\Delta S$ of two opposite holes of the first ring of holes

Figure 4 : shape birefringence of the actual fibre and of the EBHF versus wavelength

Figure 5 : Group birefringence $B_g$ due to the geometrical imperfections of the fibres computed versus the wavelength (continuous line), and group birefringence $B_G$ measured at 1540nm (■)

Figure 6 : Experimental set-up for measuring $B_G$ (in inset a typical interferogram displayed by the spectrum analyser : example of the fibre 3)

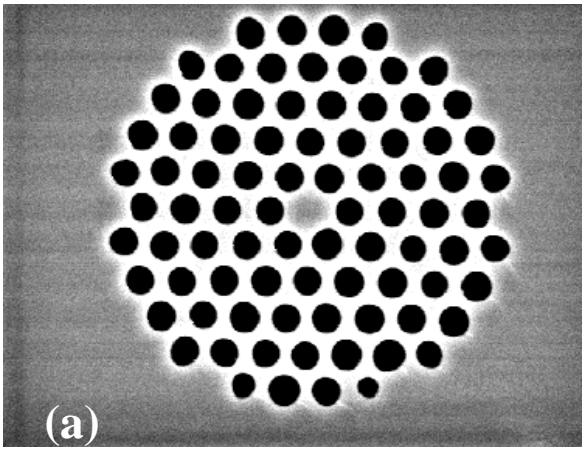
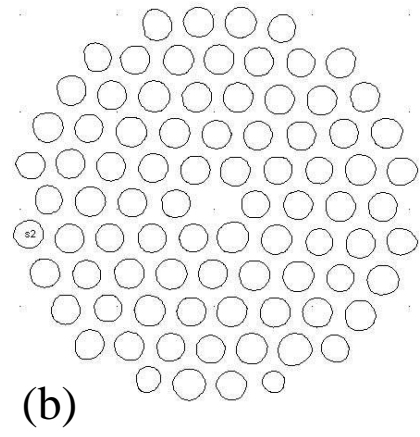
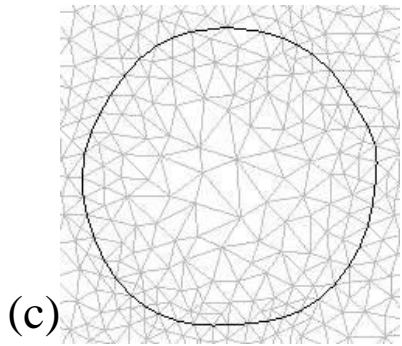
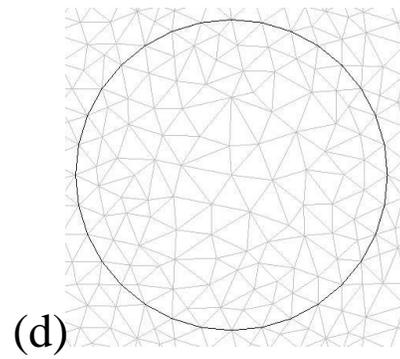

Figure 1   LABONTE

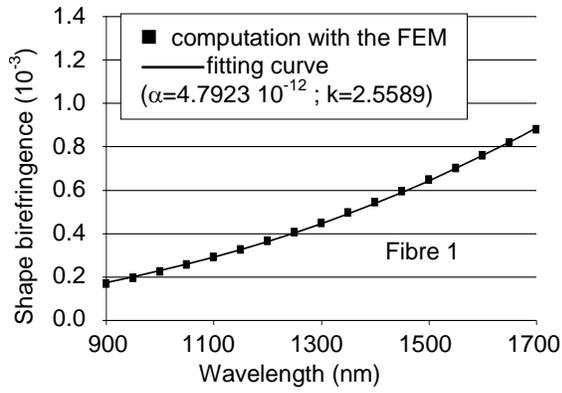
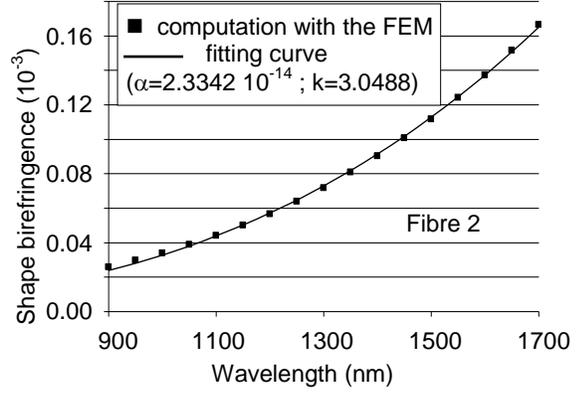
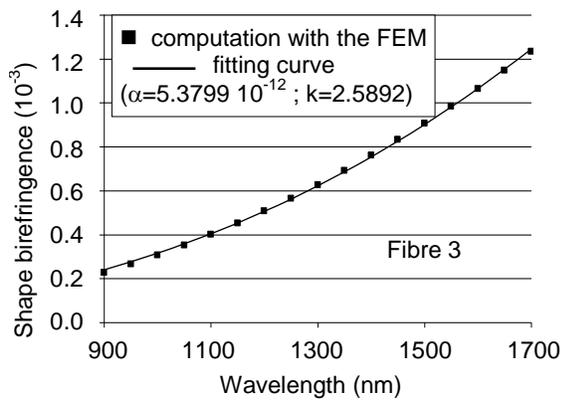
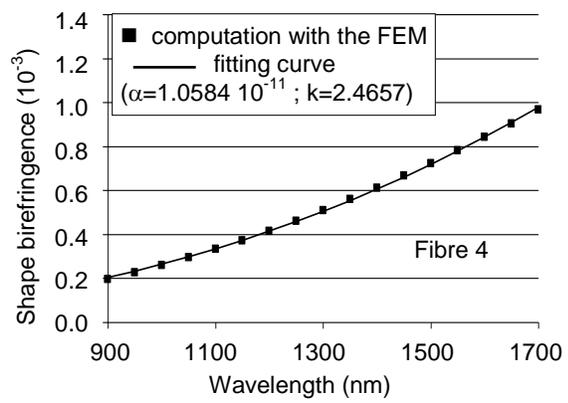

Figure 2   LABONTE

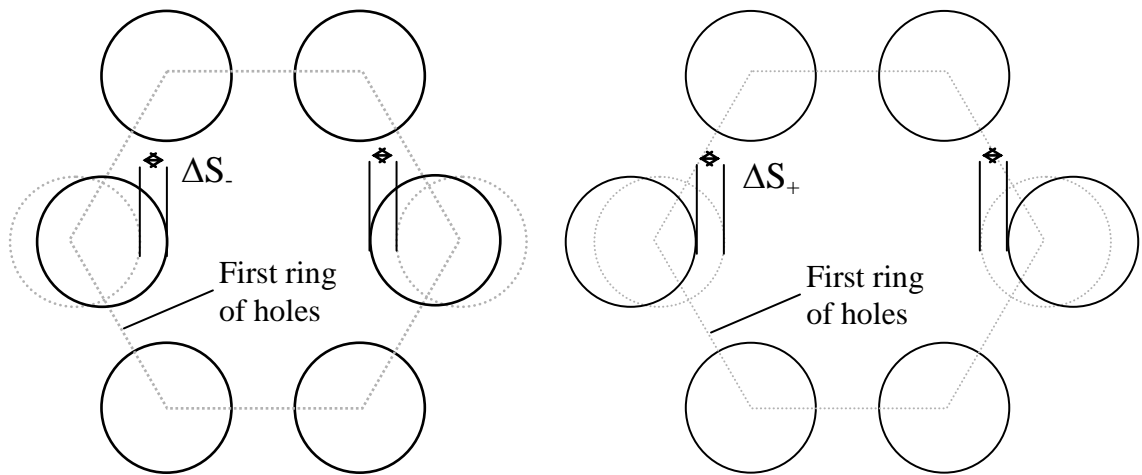

Figure 3 LABONTE

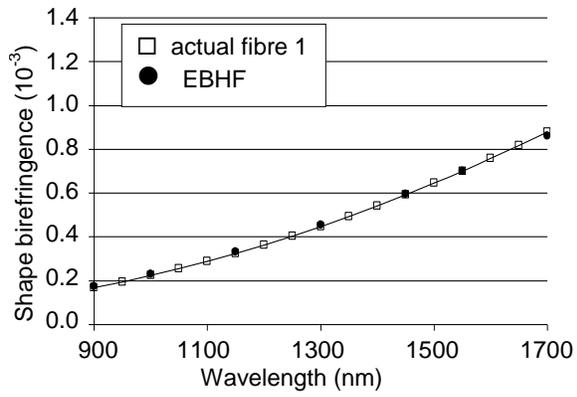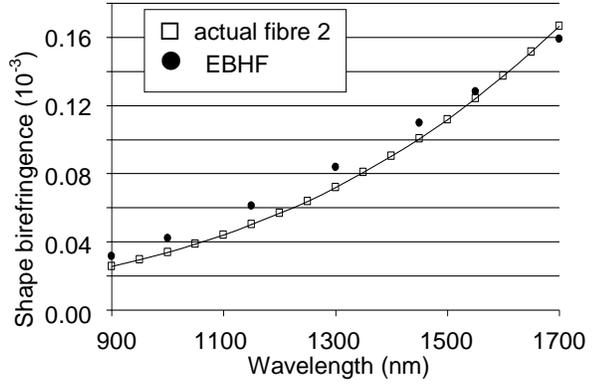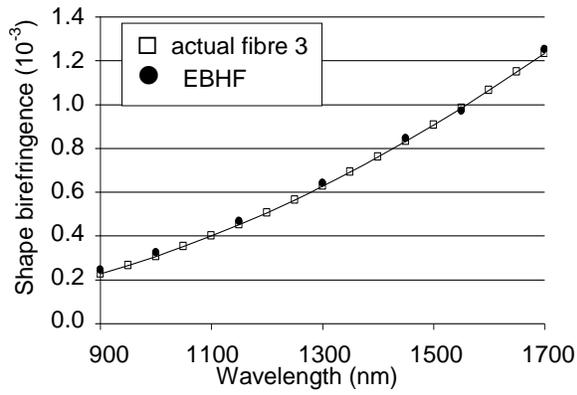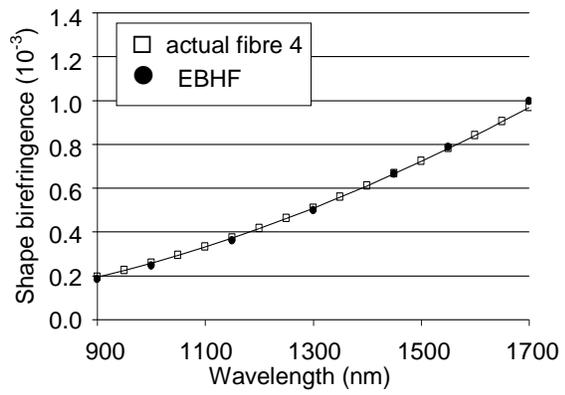

Figure 4  LABONTE

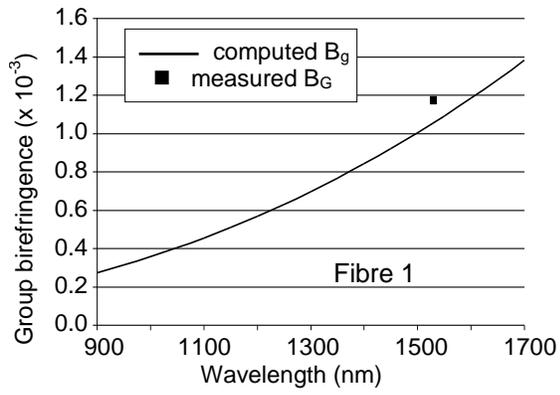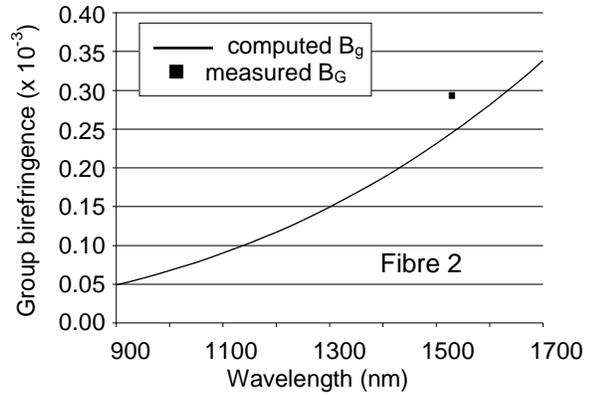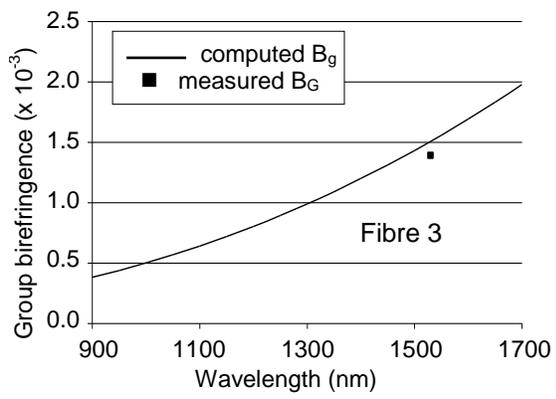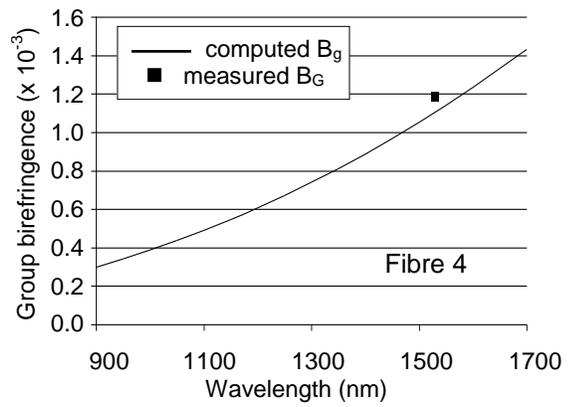

Figure 5  LABONTE

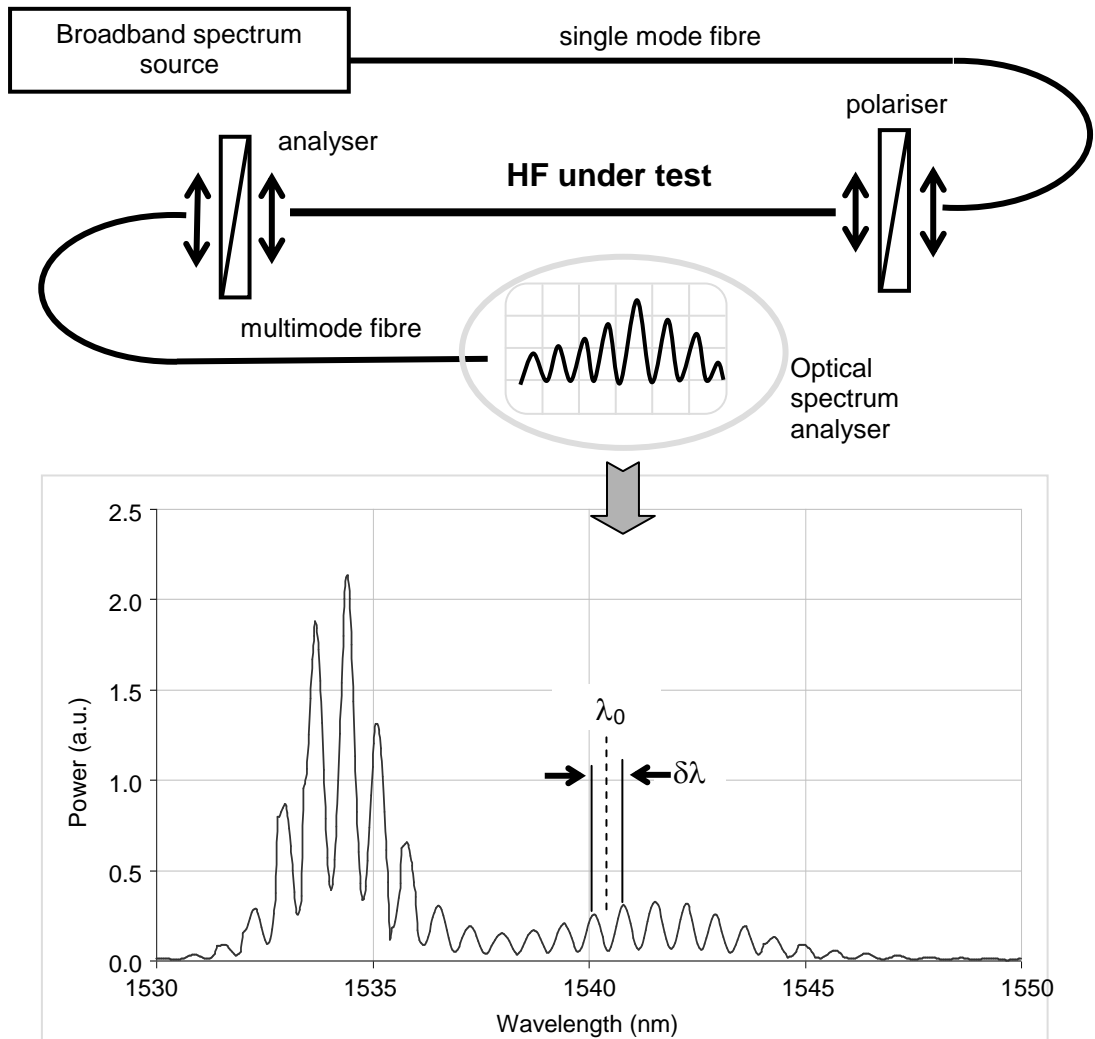

Figure 6 LABONTE